\documentstyle[a4,12pt,epsf]{article}

\newcommand{\s}{S_3}

\begin{document}

%%%%% title %%%%%
\begin{center}{\large \bf
New Approach to Texture-zeros with $S_3$ symmetry\\
\vspace{3mm}
- Flavor Symmetry and Vacuum Aligned Mass Textures -
	\footnote{talked by T. Shingai at the International Workshop on Neutrino Masses and Mixings --- Toward Unified Understanding of Quark and Lepton Mass Matrices ---, University of Shizuoka, Shizuoka, Japan, December 17-19 2006.}}
\end{center}

%%%%% author(s) %%%%%
\begin{center}
	Satoru Kaneko$^{1,}$\footnote{satoru@ific.uv.es}, Hideyuki Sawanaka$^{2,}$\footnote{hide@muse.sc.niigata-u.ac.jp}, Takaya Shingai$^{2,}$\footnote{shingai@muse.sc.niigata-u.ac.jp},

Morimitsu Tanimoto$^{3,}$\footnote{tanimoto@muse.sc.niigata-u.ac.jp} and Koichi Yoshioka$^{4,}$\footnote{yoshioka@higgs.phys.kyushu-u.ac.jp}
\vspace{6pt}\\

%%%%% address(es) %%%%%
{\it
$^1$AHEP Group, Institut de F\'{i}sica Corpuscular--C.S.I.C/Universitat de Val\`{e}ncia, \\ Edifici Instituts d'Investigaci\'{o}, Apt. 22085, E-46071 Val\`{e}ncia, Spain}

{\it
$^2$Graduate school of Science and Technology, 
Niigata University, Niigata 950-2181, Japan
}

{\it
$^3$Physics Department, Niigata University, Niigata 950-2181, Japan}

{\it
$^4$Physics Department, Kyushu University, Fukuoka 812-8581, Japan}

\end{center}

\begin{abstract}
A texture-zeros is an approach to reduce the number of free parameters in 
Yukawa couplings and it is one of the most attractive ones. In our 
paper, we discuss the origin of zero-structure in texture-zeros
by $\s$ flavor symmetry approach. Some of electroweak doublet 
Higgs fields have vanishing vacuum expectation value(VEV) which leads to 
vanishing elements in quark and lepton mass matrices. Then, the structure of 
supersymmetric scalar potential is analyzed and Higgs fields have non-trivial
$\s$ charges. As a prediction of our paper, a lower bound of a MNS matrix element, $U_{e3} \geq 0.04$, is obtained. The suppression of flavor-changing neutral 
currents(FCNC) mediated by the Higgs fields is discussed and lower bounds of the Higgs masses are derived.
\end{abstract}

\section{Introduction}
It is known that the standard model of elementary particle physics has many free Yukawa couplings. The texture-zeros was considered as an approach to reduce the number of the Yukawa couplings. By this approach, predictions among observable values are obtained. Due to useful features of the texture-zeros, many authors have studied the texture-zeros\cite{texture-zeros}-\cite{Rodejohann}. However, an origin of zero elements of mass matrices are not known. Namely, such zeros have often been assumed by hand. Our paper proposes a new approach to solve this problem and some models are considered by $U(1)$ flavor symmetry \cite{U(1) by U(1)} or discrete flavor symmetry \cite{Grimus Joshipura Lavoura and Tanimoto}. 
 
In our model, we take the discrete flavor symmetry approach. Flavor symmetry is expected to be a clue to understand the masses and mixing angles of quarks and leptons. It reduces the number of free parameters in Yukawa couplings such as the texture-zeros and some testable predictions of masses and mixing angles generally follow. Some predictive models with discrete flavor symmetries have been explored by many authors \cite{S3}-\cite{others}.We consider a $\s$ group as the discrete symmetry because $\s$ group is the simplest discrete non-Abelian group. $\s$ is a permutation group of three objects.

In this paper, we discuss the mass matrix of quarks and leptons in the case that $SU(2)$ doublet Higgs fields have non-trivial $\s$ flavor charges. Then, we discuss a dynamical realization of the texture-zeros. In previous models, some Yukawa couplings are restricted by discrete flavor symmetry and then some favored textures(texture-zeros) are obtained by changing basis. However, there is another possibility to obtain zero matrix elements. This is vanishing Higgs VEV. This possibility, however, does not make sense if a model contains only one Higgs field. Therefore, we introduce some Higgs fields. We perform the analysis of Higgs potential at the electroweak scale and examine whether some of Higgs fields in the flavor basis have vanishing expectation values. Vanishing elements(texture-zeros) of fermion mass matrices are obtained dynamically in the vacuum of the theory, and their positions are controlled by the flavor symmetry. An interesting point of the scheme is that, due to the $\s$ group structure, the up and down quark mass matrices automatically have different forms at the vacuum, which lead to non-vanishing generation mixing. 

It is found that the exact $\s$ model has Nambu-Goldstone bosons. However, these disappear by introducing $\s$ soft breaking terms and it is showed that these terms have no effect on predictions of the exact $\s$ model. Generally speaking, $\s$ soft breaking terms give rise to changing conditions necessary to take a favored vacuum. However, $\s$ soft breaking terms introduced in our model do not generate conditions for a favored vacuum. As a prediction for a unitary matrix element $U_{e3}$, we obtain $U_{e3}\geq 0.04$. This will be tested in future experiments such as the double Chooz. The suppression of FCNC mediated by multiple Higgs fields will also be discussed. 

\section{$\s$ invariant mass matrix on Supersymmetry}
In this section, the $S_3$ invariant mass matrices are presented{\cite{HY}}. We consider a supersymmetric theory and both left-handed and right-handed fermions transform under a single $\s$ symmetry. $\s$ has three irreducible representations, a singlet, a pseudo singlet and a doublet. We suppose that two of three generations belong to $S_3$ doublets and the others are singlets. Namely, $\s$ doublet is assigned to matter superfield of first and second generation,$\Psi_{L_{1,2}}$ and $\Psi_{R_{1,2}}$.
Then, the most general supersymmetric Yukawa mass terms are obtained as
\begin{equation}
	W_y = \Psi_{Li} (M_D)_{ij} \Psi_{Rj},
\end{equation}
\begin{equation}
   M_{D}=
      \left(
          \begin{tabular}{cc|c}
	      $aH_{1}$ & $b H_{S} + c H_{A}$ & $d H_{2}$ \\
	      $b H_{S} - c H_{A}$ & $a H_{2}$ & $d H_{1}$ \\
	      \hline
	      $e H_{2}$ & $e H_{1}$ & $ f H_{S}$
	        \end{tabular}
					\right),\label{Dirac mass}
\end{equation}
where $a, b, \cdots,f$ are independent Yukawa coupling constants. $H_1$ and $H_2$ are $\s$ doublet Higgs, $H_A$ is $\s$ pseudo singlet Higgs and $H_S$ is $\s$ singlet Higgs. Eq. (\ref{Dirac mass}) is obtained by using a decomposition of a tensor product of $S_3$ doublet, $\phi$ and $\psi$, which is given as
\begin{equation} 
	\begin{array}{ccccccc}
		\phi^{c} \times \psi &=& (\phi_2 \psi_2,\phi_1 \psi_1)^{T}_{\bf 2} & + & (\phi_1 \psi_2 - \phi_2 \psi_1)_{\bf 1_A} & + & (\phi_1 \psi_2 + \phi_2 \psi_1)_{\bf 1_S}, 
	\end{array}
\end{equation}
where $\phi^{c} = \sigma_{1} \phi^{*}= \sigma_{1}(\phi^{*}_{1},\phi^{*}_{2})^{T},\ \psi = (\psi_{1},\psi_{2})^{T}$.
 Now we assign ${\bf 1_{S}}$ to third generation temporarily. We reconfigure this assignment later. 
 
The $\s$ invariant bare Majorana mass matrix for matter superfield $\Psi_i$($i=1,2,3$) is given as
\begin{equation}
	W_m = \Psi_i (M_R)_{ij} \Psi_j,
\end{equation}
\begin{equation}
		M_{R}=
        \left(
	    \begin{tabular}{cc|c}
		    & $M_{1}$ &   \\
	      $M_{1}$ &   &  \\
	      \hline
	             &    &  $M_{2}$
	    \end{tabular}
	\right),
\end{equation}
where $M_1, M_{2}$ are majorana masses.
\section{$\s$ invariant Higgs scalar potential analysis}
In our model we consider eight Higgs bosons: $H_{uS}$, $H_{dS}$, $H_{uA}$, $H_{dA}$, $H_{u1}$, $H_{d1}$, $H_{u2}$ and $H_{d2}$, which are $SU(2)$ weak doublet.
Our purpose in this section is to discuss whether or not there are some vanishing VEV patterns with no parameter relations. $\s$ invariant potential is given as
\begin{equation}
  V \,=\, V_{\rm susy} + V_{\rm soft}\ ,
\end{equation}
\begin{eqnarray}
  V_{\rm susy} &=& \frac{1}{2} D^2_Y +\frac{1}{2}\sum_{a=1,2,3}(D^a)^2 
  \nonumber \\
  && +\left|\mu_S\right|^2 \left( H^\dagger_{uS}H_{uS} 
    +H^\dagger_{dS}H_{dS} \right) 
  +\left|\mu_A\right|^2 \left( H^\dagger_{uA}H_{uA} 
    +H^\dagger_{dA}H_{dA} \right) \nonumber \\
  && +\left|\mu_D\right|^2 \left( H^\dagger_{u1}H_{u1} 
    +H^\dagger_{d1}H_{d1} +H^\dagger_{u2}H_{u2} 
    +H^\dagger_{d2}H_{d2}\right), 
\end{eqnarray}
\begin{eqnarray}
  V_{\rm soft} &=& m^2_{uS}H^\dagger_{uS}H_{uS}
  +m^2_{dS} H^\dagger_{dS}H_{dS} +\left(b_S H_{uS}\epsilon H_{dS} 
    +{\rm h.c.} \right)  \nonumber \\
  && +m^2_{uA} H^\dagger_{uA}H_{uA} +m^2_{dA} H^\dagger_{dA}H_{dA} 
  +\left( b_A H_{uA}\epsilon H_{dA} +{\rm h.c.} \right) \nonumber \\
  && +m^2_{uD} (H^\dagger_{u1}H_{u1} +H^\dagger_{u2}H_{u2})
  +m^2_{dD} (H^\dagger_{d1}H_{d1} +H^\dagger_{d2}H_{d2})  \nonumber \\
  && +\left[ b_D \left( H_{u1}\epsilon H_{d2} +H_{u2}\epsilon H_{d1}
    \right) +{\rm h.c.} \right]\ ,
\end{eqnarray}
where $\epsilon$ is the antisymmetric tensor, and $b_{x}$, $m_{ux}$ and $m_{dx}$ ($x=S,A,D$) are 
the holomorphic and non-holomorphic mass parameters of supersymmetry breaking, respectively. The D-terms are explicitly given by
\begin{eqnarray}
   D_{Y} &=& -\frac{1}{2} g_{Y}\left\{ (H^{\dag}_{uS}H_{uS} -
				H^{\dag}_{dS}H_{dS}) +
				(H^{\dag}_{uA}H_{uA} -
				H^{\dag}_{dA}H_{dA}) \right. \nonumber \\
				&& +\left. (H^{\dag}_{u1}H_{u1} - H^{\dag}_{d1}H_{d1}) + (H^{\dag}_{u2}H_{u2} - H^{\dag}_{d2}H_{d2})\right\}, \label{Dy}\\
   D^{a} &=& -g_{2} \left\{ (H^{\dag}_{uS} T^{a} H_{uS} + H^{\dag}_{dS}
		     T^{a} H_{dS}) + (H^{\dag}_{uA} T^{a} H_{uA} +
		     H^{\dag}_{dA} T^{a} H_{dA}) \right. \nonumber \\
				 && +\left. (H^{\dag}_{u1} T^{a} H_{u1} + H^{\dag}_{d1} T^{a} H_{d1}) + (H^{\dag}_{u2} T^{a} H_{u2} + H^{\dag}_{d2} T^{a} H_{d2})\right\},\label{D2}
\end{eqnarray}
with $T^a$($a=1,2,3$) is the $SU(2)_L$ generators. Hence, we analyze eight equations at vacuum:
\begin{eqnarray} &&
  (\left|\mu_S\right|^2 +m^2_{uS})v_{uS} \,=\, 
  b_S v_{dS} -X v_{uS},
  (\left|\mu_S\right|^2 +m^2_{dS})v_{dS} \,=\,
  b_S v_{uS} +X v_{dS}, 
  \label{equation1} \\ &&
  (\left|\mu_A\right|^2 +m^2_{uA})v_{uA} \,=\, 
  b_A v_{dA} -X v_{uA},
  (\left|\mu_A\right|^2 +m^2_{dA})v_{dA} \,=\,
  b_A v_{uA} +X v_{dA}, 
  \label{equation2} \\ &&
  (\left|\mu_D\right|^2 +m^2_{uD})v_{u1} \,=\,
  b_D v_{d2} -X v_{u1},
  (\left|\mu_D\right|^2 +m^2_{dD})v_{d2} \,=\,
  b_D v_{u1} +X v_{d2},
  \label{equation3} \\ &&
  (\left|\mu_D\right|^2 +m^2_{uD})v_{u2} \,=\,
  b_D v_{d1} -X v_{u2},
  (\left|\mu_D\right|^2 +m^2_{dD})v_{d1} \,=\,
  b_D v_{u2} + X v_{d1},
  \label{equation4}
\end{eqnarray}
where the quantities $v_x$ are the absolute values of the neutral Higgs scalars,
\begin{eqnarray}&&
  v_{uS} \,=\, \left|\left\langle H^0_{uS}\right\rangle\right|, \quad
  v_{uA} \,=\, \left|\left\langle H^0_{uA}\right\rangle\right|, \quad
  v_{u1} \,=\, \left|\left\langle H^0_{u1}\right\rangle\right|, \quad
  v_{u2} \,=\, \left|\left\langle H^0_{u2}\right\rangle\right|, 
  \nonumber \\ &&
  v_{dS} \,=\, \left|\left\langle H^0_{dS}\right\rangle\right|, \quad
  v_{dA} \,=\, \left|\left\langle H^0_{dA}\right\rangle\right|, \quad
  v_{d1} \,=\, \left|\left\langle H^0_{d1}\right\rangle\right|, \quad
  v_{d2} \,=\, \left|\left\langle H^0_{d2}\right\rangle\right|,
\end{eqnarray}
and the parameters $b_S$, $b_A$ and $b_D$ have been chosen to be real and positive using field redefinitions. 
$X$ is defined as
\begin{equation}
	X \,\equiv\, \frac{g^2_Y + g^2_2}{4}\left( v^2_{uS}-v^2_{dS}+v^2_{uA}-v^2_{dA}+v^2_{u1}-v^2_{d1} +v^2_{u2}-v^2_{d2} \right),
\end{equation}
where $g_Y$ and $g_2$ are gauge couplings of $U(1)_Y$ and $SU(2)_L$, respectively.
\begin{table}[t]
\begin{center}
\begin{tabular}{|c|c|c|c|c|c|c|c|} \hline
$v_{uS}$ & $v_{dS}$ & $v_{uA}$ & $v_{dA}$ & $v_{u1}$ & $v_{u2}$ 
& $v_{d1}$ & $v_{d2}$ \\ \hline \hline
 $0$ & $0$ & $0$ & $0$ & $0$ &     &     & $0$ \\ \hline
 $0$ & $0$ & $0$ & $0$ &     & $0$ & $0$ &     \\ \hline
 $0$ & $0$ & $0$ & $0$ &     &     &     &     \\ \hline
 $0$ & $0$ &     &     & $0$ & $0$ & $0$ & $0$ \\ \hline
 $0$ & $0$ &     &     & $0$ &     &     & $0$ \\ \hline
 $0$ & $0$ &     &     &     & $0$ & $0$ &     \\ \hline
 $0$ & $0$ &     &     &     &     &     &     \\ \hline
     &     & $0$ & $0$ & $0$ & $0$ & $0$ & $0$ \\ \hline
     &     & $0$ & $0$ & $0$ &     &     & $0$ \\ \hline
     &     & $0$ & $0$ &     & $0$ & $0$ &     \\ \hline
     &     & $0$ & $0$ &     &     &     &     \\ \hline
     &     &     &     & $0$ & $0$ & $0$ & $0$ \\ \hline
     &     &     &     & $0$ &     &     & $0$ \\ \hline
     &     &     &     &     & $0$ & $0$ &     \\ \hline
\end{tabular}
\caption{All possible minima of the scalar potential for $S_3$ singlet
and doublet Higgs fields without tuning of Lagrangian parameters for
electroweak symmetry breaking. The blank entries denote
non-vanishing VEVs.}
{\label{summary}}
\end{center}
\end{table} 
In general, these equations are the coupled equations through a common parameter $X$ which contains all the Higgs VEVs. However, we can separate these equations into three parts for the singlet, the pseudo singlet and the doublet. This is because vanishing-VEVs makes the equations trivial within each sector. 
Consequently, possible 14 VEV patterns in Table {\ref{summary}} are obtained. They do not need some parameter conditions. 

\section{Quark and lepton mass textures}
We obtained 14 VEV patterns in previous section.
Now let us analyze these patterns phenomenologically. The most interesting pattern of 14 patterns is the following:
\begin{equation}
	v_{uS}=v_{dS}=v_{u1}=v_{d2}=0,\qquad v_{uA},v_{dA},v_{u2},v_{d1}\not= 0. {\label{vev}}
\end{equation}
This pattern leads to the simplest texture(i.e.the maximal number of zero matrix elements) with non-trivial flavor mixing. 

We consider mass matrices obtained from the VEV pattern (\ref{vev}). In section 2, we assigned $\s$ singlet ${\bf	1_S}$ to third generation. However, we can assign ${\bf 1_S}$ to any generation in general. For example, ${\bf 1_S}$ can be also assigned to first generation. As results of exhausting all the $S_3$ charge assignments, mass matrices are derived as 
\begin{equation}
	M_u =
		\left(
			\begin{array}{ccc}
					& b_u &		\\
				d_u &		& f_u \\
				 & - f_u & i_u
			\end{array}
		\right),\qquad
	M_d =
		\left(
			\begin{array}{ccc}
					& b_d &		\\
				d_d &	e_d	&  \\
				-e_d &  & i_d
			\end{array}
		\right), \qquad
	M_e =
		\left(
			\begin{array}{ccc}
					& d_d &	3e_d	\\
				b_d &	-3 e_d	& \\
				 &  & i_d
			\end{array}
			\right),\label{quark charged lepton}
\end{equation}
\begin{equation}
	M_{\nu} =
		\left(
			\begin{array}{ccc}
				& b_{\nu} &	c_{\nu}	\\
				-b_{\nu} &	e_{\nu}	&  \\
				g_{\nu} & & 
			\end{array}
		\right),\qquad
	M_R =
		\left(
			\begin{array}{ccc}
					& M_1 &		\\
				M_1 &		&  \\
				 &  & M_2
			\end{array}
			\right),\label{neutrino}
\end{equation}
where blank entries denote zero and each parameter in $M_{u},M_{d},M_{e},M_{\nu}$ such as $d_{u},d_{d},b_{\nu}$ stands for a product of a Yukawa coupling and a VEV, for example, $d_{u} = d v_{u2}$. We assume here the $SU(5)$ grand unification. Namely, $\s$ charge assignment of (\ref{quark charged lepton}) and (\ref{neutrino}) is embedded in the $SU(5)$ grand unification. It is, however, noted that the charge assignment of the right-handed neutrinos is completely irrelevant to low-energy physics, and the generation structure of the light neutrino mass matrix is determined only by the flavor charge of the left-handed leptons. We have included a group theoretical factor $-3$ \cite{GJ} in front of the element $e_d$. Such a factor originates from a Yukawa coupling to higher-dimensional Higgs field. For example, this originates from 45-plet of $SU(5)$.

Some predictions from our model are given as
\begin{equation}
	\left| V_{cb} \right| = \sqrt{\frac{m_c}{m_t}},\qquad \left| U_{e3} \right| \geq 0.04.\label{prediction}
\end{equation}
The former prediction in (\ref{prediction}) is known to accurately fit the experimental data and it has already been studied in other theoretical framework \cite{WY,Vcb}. A precise numerical estimation, however, indicates that the former prediction is not satisfied with the experimental data slightly. There is a possibility to modify this deviation. That is a supersymmetric threshold correction. The theory would become viable in light of the current experimental data. We leave a detailed discussion of this correction to future investigations. The latter prediction\cite{PDG} in (\ref{prediction}) is most important prediction in our model. This lower bound will be tested in future experiments, such as the double Chooz\cite{doublechooz}.

\section{Higgs mass spectrum and $\s$ soft breaking terms in B-term}

The $S_3$ potential has an enhanced global symmetry $SU(2) \times U(1)^2$  and leads to massless Nambu-Goldstone bosons 
in the electroweak broken phases. Global symmetries $U(1)^2$ is presented in Table \ref{U(1)}.
\begin{table}[t]
	\begin{center}
		\caption{The $U(1)$ symmetries of the Higgs scalar potential with the $\s$ breaking terms (\ref{soft breaking terms}).\label{U(1)}}
     \begin{tabular}{|c|c|c|c|c|c|c|c|c|}
			 \hline
			 & $H_{uS}$ & $H_{dS}$ & $H_{uA}$ & $H_{dA}$ & $H_{u1}$ & $H_{u2}$ & $H_{d1}$ & $H_{d2}$ \\
			 \hline \hline
			 $U_X$ & $+1$ & $-1$ & $0$ & $0$ & $+1$ & $0$ & $0$ & $-1$ \\
			 \hline
			 $U_Y$ & $+1$ & $-1$ & $+1$ & $-1$ & $+1$ & $+1$ & $-1$ & $-1$\\
			 \hline
		\end{tabular}
	\end{center}
\end{table}
It is therefore reasonable to softly break the flavor symmetry within the scalar potential.
We introduce the following supersymmetry-breaking soft terms which do not break phenomenological characters of the exact $S_3$ model.
\begin{equation}
	V_{\not S_3} = b_{SD}H_{uS}H_{d2}+ b'_{SD}H_{u1}H_{dS}+ b_{AD}H_{uA}H_{d1}+ b'_{AD}H_{u2}H_{dA} + {\rm h.c.}\label{soft breaking terms}
\end{equation}
A $\s$ soft breaking model have the same phenomenological characters as the exact $S_3$ model and we can take the VEV pattern ({\ref{vev}}) with no parameter condition.

The $\s$ soft breaking model also predicts a lightest Higgs mode. This is similar to the minimal supersymmetric standard model(MSSM). We obtain the neutral Higgs masses squared approximately:
\begin{eqnarray}
  &&\{M^2_{h^0},\ M^2_{H^0_1},\ M^2_{H^0_2},\ M^2_{H^0_3},\ M^2_{H^0_4}\} \nonumber\\
  &&\qquad = \{ {\cal O}(v^2),\ \bar m^2-\bar b -\bar b',\ 
  \bar m^2 +\bar b -\bar b',\ \bar m^2 -\bar b +\bar b',\ 
  \bar m^2 +\bar b +\bar b'\}.
\end{eqnarray}
The other three mass eigenvalues squared take the complicated forms. Then, we 
 make the simplifying assumption
\begin{eqnarray}
  & \left|\mu_x\right|^2 +m^2_{ux} \,=\, 
  \left|\mu_x\right|^2 +m^2_{dx} \,\equiv\, \bar m^2, \qquad
  b_x \,\equiv\, \bar b, \qquad  (x=S,A,D) \\
  & b_{SD}=b'_{SD}=b_{AD}=b'_{AD} \,\equiv\, \bar b'\ ,
\end{eqnarray}
with the hierarchy ${\bar m}^2,\,|\bar b|,\,|\bar b'|\,\gg\,v^2
\equiv v^2_{u2}+v^2_{d1}+v^2_{uA}+v^2_{dA}$. Hence, we obtain a lightest mode 
which has a weak scale mass ${\cal O}(v)$ and its eigenvector is given as
\begin{equation}
  h^0 \,=\, \frac{1}{v}\left( v_{uA} h^0_{uA} +v_{dA} h^0_{dA} 
    +v_{u2}h^0_{u2} +v_{d1}h^0_{d1} \right).
\end{equation}

\begin{figure}[t]
	\begin{center}
		\epsfxsize=9.0cm
\centerline{\epsfbox{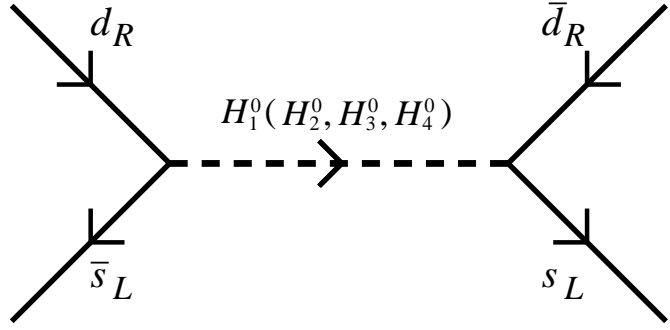}}
\caption{A Higgs-mediated tree-level FCNC process for the K meson system\label{tree level FCNC}}
\end{center}
\end{figure}

\section{$K^0$ - ${\bar K^0}$ mixing}
\begin{figure}[t]
	\begin{center}
		\epsfxsize=9.0cm
\centerline{\epsfbox{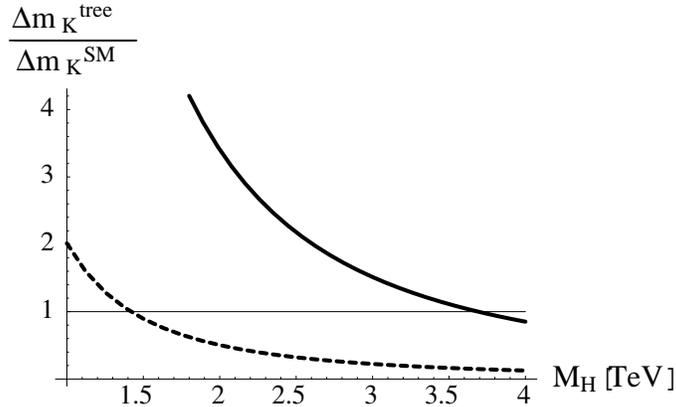}}
\caption{$\Delta m^{{\rm tree}}_K / \Delta m^{{\rm SM}}_K $ as a function of the Higgs parameter $M_H$. The solid and dashed carves correspond to $\eta = 0$ and $0.03$, respectively.\label{ratio}}
\end{center}
\end{figure}

Since there are multiple electroweak doublet Higgs bosons which couple to 
matter fields, flavor-changing processes are mediated at classical 
level by these Higgs fields.	We can show that flavor-changing process at classical level leads to all but one Higgs masses which are order of supersymmetric breaking scale. Therefore, the experimental observations of FCNC rare events would show a bound on the supersymmetry breaking scale.
Among various experimental constraints, we find the most important constraint comes from the neutral K meson mixing. For the heavy Higgs modes, the tree-level $K_{L}-K_{S}$ mass difference $\Delta m^{\rm tree}_K$ is given by the matrix element of the effective Hamiltonian between K mesons{\cite{FCNC}}.
$\Delta m^{\rm tree}_K$ includes $M_{H}$, which is an average of the Higgs masses $1/M^{2}_{H} = \frac{1}{4}\left( 1/M^{2}_{H^{0}_{1}} + 1/ M^{2}_{H^{0}_{2}} + 1/M^{2}_{H^{0}_{3}}+1/M^{2}_{H^{0}_{4}} \right)$, and a free parameter $\eta$ is given as
\begin{equation}
	\eta = \frac{(y^S_d)_{22} b_d v_{d1}}{m^2_b} - \frac{(y^S_d)_{13} d_d v_{d1}}{m_s m_b}.
\end{equation}
$(y^S_d)_{22}$ and $(y^S_d)_{13}$ are the down type quark Yukawa couplings. It is found from Fig. \ref{ratio} that heavy Higgs masses are bounded from below so as to suppress the extra Higgs contribution compared with the standard model one. $M_H$ is roughly given by
\begin{equation}
  M_{H} \ge
	    \left\{
			    \begin{array}{cl}
						3.8 {\rm TeV} &(\eta = 0) \\
						1.4 {\rm TeV} &(\eta = 0.03) \\
					\end{array}
				\right.
,
\end{equation}
where we took $\eta = 0$ and $0.03$ as typical values. %and $\eta$ could not become larger because order of vacuums is weak scale. 
We used the experimental values $m_K=490 {\rm MeV}$ and $f_K=160 {\rm MeV}$, and took $v_{d1}=100{\rm GeV}$ as a typical electroweak scale.

\section{Summary}
The mass matrix of quarks and leptons have discussed in view of $\s$ flavor symmetry approach. We have analyzed the structure of the Higgs scalar potential and examined possible vanishing elements of quark and lepton mass matrices. Then, there exists the most favored VEV pattern for which parameter conditions are not necessary and quark and lepton mass matrices from the VEV pattern is consistent with experimental data. $\s$ assignment to the flavor can be embedded in the $SU(5)$ grand unification and we take non-trivial $\s$ assignment. As phenomenological results, a prediction for the lepton mixing $U_{e3}$ is obtained: $U_{e3} \geq 0.04$. This prediction will be tested in future experiments. It is found from tree-level FCNC processes mediated by the Higgs bosons that TeV scale typical Higgs mass is consistent with $K^0$-${\bar K^0}$ mixing experimental bound.

\section*{Acknowledgements}
We would like to thank the organizers of the International Workshop on Neutrino Masses and Mixings and thank J.~Kubo, T.~Kobayashi and H.~Nakano for helpful discussions.
\appendix


\begin{thebibliography}{0}
		\bibitem{texture-zeros}
P.~H.~Frampton, S.~L.~Glashow and T.~Yanagida,
Phys. Lett. {\bf B548} (2002), 119.
\\
M.~Raidal and A.~Strumia, 
Phys. Lett. {\bf B553} (2003), 72.
\\
Z.~Xing, 
Phys. Rev. {\bf D69} (2004), 013006.
\\
G.~C.~Branco, R.~Gonzalez Felipe, F.~R.~Joaquim and T.~Yanagida,
Phys. Lett. {\bf B562} (2003), 265.

%%%%%%%%%%%%%%%%%%%%%%%%%%%%%%%%%%%%%%%%%%%
\bibitem{F3R}  
%%%%%%%%%%%%%%%%%%%%%%%%%%%%%%%%%%%%%%%%%%%
H.~Fritzsch,
Phys. Lett. {\bf B73} (1978), 317;
Nucl. Phys. {\bf B155} (1979), 189.
\\
P.~Ramond, R.~G.~Roberts and G.~G.~Ross, 
Nucl. Phys. {\bf B406} (1993), 19.

%%%%%%%%%%%%%%%%%%%%%%%%%%%%%%%%%%%%%%%%%%%
\bibitem{pre}  
%%%%%%%%%%%%%%%%%%%%%%%%%%%%%%%%%%%%%%%%%%%
H.~Nishiura, K.~Matsuda and T.~Fukuyama,
Phys. Rev. {\bf D60} (1999), 013006.
\\
E.~K.~Akhmedov, G.~C.~Branco and M.~N.~Rebelo,
Phys. Rev. Lett. {\bf 84} (2000), 3535.
\\
S.~K.~Kang and C.~S.~Kim,
Phys. Rev. {\bf D63} (2001), 113010.

%%%%%%%%%%%%%%%%%%%%%%%%%%%%%%%%%%%%%%%%%%%
\bibitem{neutrino}
%%%%%%%%%%%%%%%%%%%%%%%%%%%%%%%%%%%%%%%%%%%
P.~H.~Frampton, S.~L.~Glashow and D.~Marfatia,
Phys. Lett. {\bf B536} (2002), 79.
\\
Z.-z.~Xing,
Phys. Lett. {\bf B530} (2002), 159.
\\
R.~Barbieri, T.~Hambye and A.~Romanino,
J. High Energy Phys. {\bf 03} (2003), 017.
\\
A.~Ibarra and G.~G.~Ross,
Phys. Lett. {\bf B575} (2003), 279.

%%%%%%%%%%%%%%%%%%%%%%%%%%%%%%%%%%%%%%%%%%%
\bibitem{tanimoto}
%%%%%%%%%%%%%%%%%%%%%%%%%%%%%%%%%%%%%%%%%%%
A.~Kageyama, S.~Kaneko, N.~Shimoyama and M.~Tanimoto,
Phys. Lett. {\bf B538} (2002), 96.
\\
S.~Kaneko and M.~Tanimoto,
Phys. Lett. {\bf B551} (2003), 127.
\\
M.~Honda, S.~Kaneko and M.~Tanimoto,
J. High Energy Phys. {\bf 09} (2003), 028;
Phys. Lett. {\bf B593} (2004), 165.
\\
S.~Kaneko, H.~Sawanaka and M.~Tanimoto,
J. High Energy Phys. {\bf 08} (2005), 073.

%%%%%%%%%%%%%%%%%%%%%%%%%%%%%%%%%%%%%%%%%%%
\bibitem{BO}
%%%%%%%%%%%%%%%%%%%%%%%%%%%%%%%%%%%%%%%%%%%
M.-C.~Chen and K.T.~Mahanthappa,
Phys. Rev. {\bf D68} (2003), 017301.
\\
M.~Bando and M.~Obara,
Prog. Theor. Phys. {\bf 109} (2003), 995.
\\
M.~Bando, S.~Kaneko, M.~Obara and M.~Tanimoto,
Phys. Lett. {\bf B580} (2004), 229.

%%%%%%%%%%%%%%%%%%%%%%%%%%%%%%%%%%%%%%%%%%%
\bibitem{inverted}
%%%%%%%%%%%%%%%%%%%%%%%%%%%%%%%%%%%%%%%%%%%
L.~Lavoura,
Phys. Lett. {\bf B609} (2005), 317.
\\
W.~Grimus, S.~Kaneko, L.~Lavoura, H.~Sawanaka and M.~Tanimoto, 
J. High Energy Phys. {\bf 01} (2006), 110.

%%%%%%%%%%%%%%%%%%%%%%%%%%%%%%%%%%%%%%%%%%%
\bibitem{FS}
%%%%%%%%%%%%%%%%%%%%%%%%%%%%%%%%%%%%%%%%%%%
M.~Frigerio and A.~Y.~Smirnov,
Phyes. Rev. {\bf D67} (2003), 013007.

%%%%%%%%%%%%%%%%%%%%%%%%%%%%%%%%%%%%%%%%%%%
\bibitem{WY}
%%%%%%%%%%%%%%%%%%%%%%%%%%%%%%%%%%%%%%%%%%%
N.~Uekusa, A.~Watanabe and K.~Yoshioka, 
Phys. Rev. {\bf D71} (2005), 094024.
\\
A.~Watanabe and K.~Yoshioka,
J. High Energy Phys. {\bf 05} (2006), 044.

%%%%%%%%%%%%%%%%%%%%%%%%%%%%%%%%%%
\bibitem{Rodejohann}
C.~Hagedorn and W.~Rodejohann,
J. High Energy Phys. {\bf 07} (2005), 034.
\\
A.~Merle and W.~Rodejohann,
Phys. Rev. {\bf D73} (2006), 073012.

\bibitem{U(1) by U(1)}
L.~Ib\'{a}\~{n}ez and G.~G.~Ross, 
Phys. Lett. {\bf B332} (1994), 100.
\\
P.~Bin\'{e}truy, S.~Lavignac and P.~Ramond, 
Phys. Lett. {\bf B350} (1995), 49; Nucl. Phys. {\bf B477} (1996), 353.
\\
P.~Bin\'{e}truy, S.~Lavignac, S.~T.~Petcov and P.~Ramond, 
Nucl. Phys. {\bf B496} (1997), 3.
\\
J.~K.~Elwood, N.~Irges and P.~Ramond, 
Phys. Rev. Lett. {\bf 81} (1998), 5064.
\\
N.~Irges, S.~Lavignac and P.~Ramond, 
Phys. Rev. {\bf D58} (1998), 035003.
\bibitem{Grimus Joshipura Lavoura and Tanimoto}
For example, W.~Grimus, A.~S.~Joshipura, L.~Lavoura and M.~Tanimoto, 
Eur. Phys. J. {\bf C36} (2004), 227.

\bibitem{S3}
S.~Pakvasa and H.~Sugawara,
 Phys. Lett. B \textbf{73} (1978), 61.
\\
D.~Wyler,
 Phys. Rev. D \textbf{19} (1979), 330.
\\
L.~J.~Hall and H.~Murayama,
 Phys. Rev. Lett. \textbf{75} (1995), 3985.
\\
C.~D.~Carone, L.~J.~Hall and H.~Murayama, 
%``$(S_3)~3$ flavor symmetry and $p\to K~0 e~+$,''
Phys. Rev. D \textbf{53}, (1996) 6282. 
\\
R.~Dermisek and S.~Raby,
 Phys. Rev. D \textbf{62} (2000), 015007.
\\
R.~N.~Mohapatra, A.~Perez-Lorenzana and C.~A.~de Sousa Pires,
 Phys. Lett. B \textbf{474} (2000), 355.
\\
J.~Kubo, A.~Mondragon, M.~Mondragon and E.~Rodriguez-Jauregui,
 Prog. Theor. Phys. \textbf{109} (2003), 795
[Errata;\ {\bf 114} (2005), 287].
\\
T.~Kobayashi, J.~Kubo and H.~Terao,
 Phys. Lett. B \textbf{568} (2003), 83.
\\
P.~F.~Harrison and W.~G.~Scott,
 Phys. Lett. B \textbf{557} (2003), 76.
\\
J.~Kubo, H.~Okada and F.~Sakamaki,
 Phys. Rev. D \textbf{70} (2004), 036007.
\\
S.-L.~Chen, M.~Frigerio and E.~Ma,
 Phys. Rev. D \textbf{70} (2004), 073008
[Errata; D {\bf 70} (2004), 079905].
\\
W.~Grimus and L.~Lavoura, 
 J. High Energy Phys. \textbf{08} (2005), 013.
\\
Y.~Koide,  Phys. Rev. D \textbf{73} (2006), 057901.

\bibitem{HY}
N.~Haba and K.~Yoshioka,
 Nucl. Phys. B \textbf{739} (2006), 254.

\bibitem{Sn}
S.~Pakvasa and H.~Sugawara, 
Phys. Lett. B \textbf{82} (1979), 105.
\\
Y.~Yamanaka, H.~Sugawara and S.~Pakvasa,
 Phys. Rev. D \textbf{25} (1982), 1895 [Errata; D {\bf 29} (1984),
2135].
\\
C.~Hagedorn, M.~Lindner and R.~N.~Mohapatra, 
 J. High Energy Phys. \textbf{06} (2006), 042.

\bibitem{D4}
W.~Grimus and L.~Lavoura,
 Phys. Lett. B \textbf{572} (2003), 189.
\\
W.~Grimus, A.~S.~Joshipura, S.~Kaneko, L.~Lavoura and M.~Tanimoto,
 J. High Energy Phys. \textbf{07} (2004), 078.
\\
W.~Grimus, A.~S.~Joshipura, S.~Kaneko, L.~Lavoura, H.~Sawanaka 
and M.~Tanimoto,
 Nucl. Phys. B \textbf{713} (2005), 151.
 
\bibitem{Q6}
K.~S.~Babu and J.~Kubo,  Phys. Rev. D \textbf{71} (2005), 056006.
\\
Y.~Kajiyama, E.~Itou and J.~Kubo,
 Nucl. Phys. B \textbf{743} (2006), 74.

\bibitem{Q8}
M.~Frigerio, S.~Kaneko, E.~Ma and M.~Tanimoto,
 Phys. Rev. D \textbf{71} (2005), 011901.

\bibitem{A4}
E.~Ma and G.~Rajasekaran,
 Phys. Rev. D \textbf{64} (2001), 113012.
\\
E.~Ma,
Mod.\ Phys.\ Lett.\ A {\bf 17} (2002), 2361.
\\
K.~S.~Babu, E.~Ma and J.~W.~F.~Valle,
 Phys. Lett. B \textbf{552} (2003), 207.
\\
M.~Hirsch, J.~C.~Romao, S.~Skadhauge, J.~W.~F.~Valle and A.~Villanova del
Moral,
 Phys. Rev. D \textbf{69} (2004), 093006.
\\
A.~Zee,
 Phys. Lett. B \textbf{630} (2005), 58.
\\
G.~Altarelli and F.~Feruglio,
 Nucl. Phys. B \textbf{720} (2005), 64.
\\
X.-G.~He, Y.-Y.~Keum and R.R.~Volkas,
 J. High Energy Phys. \textbf{04} (2006), 039.
\\
E.~Ma, H.~Sawanaka and M.~Tanimoto, hep-ph/0606103.

\bibitem{others}
E.~Derman and H.~S.~Tsao,
 Phys. Rev. D \textbf{20} (1979), 1207.
\\
D.~Chang, W.~Y.~Keung and G.~Senjanovic,
 Phys. Rev. D \textbf{42} (1990), 1599.
\\
D.~B.~Kaplan and M.~Schmaltz,
 Phys. Rev. D \textbf{49} (1994), 3741.
\\
P.~H.~Frampton and T.~W.~Kephart,
Int.\ J.\ Mod.\ Phys.\ A {\bf 10} (1995), 4689.
\\
A.~Aranda, C.~D.~Carone and R.~F.~Lebed,
 Phys. Lett. B \textbf{474} (2000), 170.
\\
N.~Haba, A.~Watanabe and K.~Yoshioka,
 Phys. Rev. Lett. \textbf{97} (2006), 041601.
\\
C.~Hagedorn, M.~Lindner and F.~Plentinger,
 Phys. Rev. D \textbf{74} (2006), 025007.

\bibitem{GJ} 
H.~Georgi and C.~Jarlskog,
 Phys. Lett. B \textbf{86} (1979), 297.
\\
S.~Dimopoulos, L.~J.~Hall and S.~Raby,
 Phys. Rev. Lett. \textbf{68} (1992), 1984.

\bibitem{Vcb}
K.~S.~Babu and Q.~Shafi,
 Phys. Rev. D \textbf{47} (1993), 11.
\\
K.~S.~Babu and R.~N.~Mohapatra,
 Phys. Rev. Lett. \textbf{74} (1995), 2418.

\bibitem{PDG}
S.~Eidelman et al. [Particle Data Group Collaboration],
 Phys. Lett. B \textbf{592} (2004), 1.

\bibitem{doublechooz}
K.~Anderson et al., hep-ex/0402041.

\bibitem{FCNC}
F.~Gabbiani, E.~Gabrielli, A.~Masiero and L.~Silvestrini, 
 Nucl. Phys. B \textbf{477} (1996), 321.

\end{thebibliography}
\end{document}